\documentclass[floatfix,rmp,twocolumn,twoside]{revtex4}
\usepackage{graphicx}

\bibpunct{[}{]}{,}{n}{}{,}

\usepackage{amsmath}
\usepackage[english]{babel}
\usepackage{verbatim}
\usepackage{graphicx}
\usepackage{alltt}
\usepackage[small]{caption}
\usepackage[english]{babel}
\usepackage{multirow}
\usepackage{varwidth}

\makeatletter
\g@addto@macro\@verbatim\small
\makeatother

\begin{document}

\title{A DSL for Mapping Abstract Syntax Models \\ to Concrete Syntax Models in ModelCC}
\author{Luis~Quesada, Fernando~Berzal, and Juan-Carlos~Cubero\\
  Department of Computer Science and Artificial Intelligence, CITIC, University of Granada, \\
  Granada 18071, Spain \\
  \textit{lquesada@decsai.ugr.es, fberzal@decsai.ugr.es, jc.cubero@decsai.ugr.es}
  }
%\toctitle{Lecture Notes in Computer Science}
%\tocauthor{Authors' Instructions}

\begin{abstract}
ModelCC is a model-based parser generator that decouples language design from language processing.
ModelCC provides two different mechanisms to specify the mapping from an abstract syntax model to a concrete syntax model: metadata annotations defined on top of the abstract syntax model specification and a domain-specific language for defining ASM-CSM mappings.
Using a domain-specific language to specify the mapping from abstract to concrete syntax models allows the definition of multiple concrete syntax models for the same abstract syntax model.
In this paper, we describe the ModelCC domain-specific language for abstract syntax model to concrete syntax model mappings and we showcase its capabilities by providing a meta-definition of that domain-specific language.
\end{abstract}

\maketitle
\section{Introduction} \label{sec:introduction}

\noindent Model-based language specification techniques \cite{Kleppe2007} decouple language design from language processing and automatically generate the corresponding language grammar, thus making the language design process less arduous.

ModelCC \cite{Quesada2011c,Quesada2012k} is a model-based parser generator that allows the specification of the abstract syntax elements.a language as a set of classes, which represent language elements, and relationships between those classes or language elements.

ModelCC allows mapping the abstract syntax model to concrete syntax models by imposing constraints over language elements and th(i.e. sort of a s using either metadata annotations or a domain specific language for the specification of language constraints (i.e. sort of a metalanguage).

In this paper, we propose the ModelCC domain-specific language for abstract syntax model to concrete syntax model mappings (from now on referred as the ModelCC DSL for ASM-CSM mappings) and present its specification in a model-based way using ModelCC.
This domain-specific language ultimately allows model-based parser generators to decouple abstract syntax models from concrete syntax models.

Section \ref{sec:background} introduces model-based language specification and the ModelCC model-based parser generator.
Section \ref{sec:syntaxmodelconstraintspecificationlanguage} describes ModelCC the ModelCC domain-specific language for ASM-CSM mappings.
Finally, Section \ref{sec:conclusionsandfuturework} presents our conclusions and future work.

\section{Model-Based Language Specification} \label{sec:background}

\noindent Most existing language specification techniques \cite{Aho2006} require the language designer to provide a textual specification of the language grammar. The proper specification of such a grammar is a nontrivial process that depends on the lexical and syntax analysis techniques to be used, since each kind of technique requires the grammar to comply with a specific set of constraints. Each analysis technique is characterized by its expression power and this expression power determines whether a given analysis technique is suitable for a particular language. The most significant constraints on formal language specification originate from the need to consider context-sensitivity, the need to perform an efficient analysis, and some techniques' inability to resolve conflicts caused by grammar ambiguities.

In practice, when we want to build a complex data structure from an input codified using a specific syntax, the implementation of the mandatory language processor requires the software engineer to build a grammar-based language specification for the input data and also to implement the conversion from the parse tree returned by the parser to the desired data structure, which is an instance of the data model that describes the input data.

Whenever the language specification has to be modified, the language designer has to manually propagate changes throughout the entire language processor tool chain, from the specification of the grammar defining the formal language (and its adaptation to specific parsing tools) to the corresponding data model. These updates are time-consuming, tedious, and error-prone. By making such changes labor-intensive, the traditional language processing approach hampers the maintainability and evolution of the language used to represent the data \cite{Kats2010}.

Moreover, it is not uncommon for different applications to use the same language. For example, the compiler, different code generators, and other tools within an IDE, such as the editor or the debugger, typically need to grapple with the full syntax of a programming language. Unfortunately, their maintenance typically requires keeping several copies of the same language specification in sync.

The idea behind model-based language specification is that, starting from a single abstract syntax model (ASM) that represents the core concepts in a language, language designers can develop one or several concrete syntax models (CSMs). These CSMs can suit the specific needs of the desired textual or graphical representation. The ASM-CSM mappings can be performed, for instance, by annotating the abstract syntax model with the constraints needed to transform the elements in the abstract syntax into their concrete representation.

This way, the ASM representing the language can be modified as needed without having to worry about the language processor and the peculiarities of the chosen parsing technique, since the corresponding language processor will be automatically updated. In this case, the language designer does not have to manually propagate changes throughout the language processor tool chain. Also, when different applications use the same language, there is no need to keep or maintain duplicate language models.

Finally, as the ASM is not bound to a particular parsing technique, evaluating alternative and/or complementary parsing techniques is possible without having to propagate their constraints into the language model. Therefore, by using an ASM, model-based language specification completely decouples language specification from language processing, which can be performed using whichever parsing techniques are suitable for the formal language implicitly defined by the abstract model and its concrete mapping.

\begin{figure}[tb]
\begin{minipage}[tb]{\linewidth}
\centering
\includegraphics[scale=0.225]{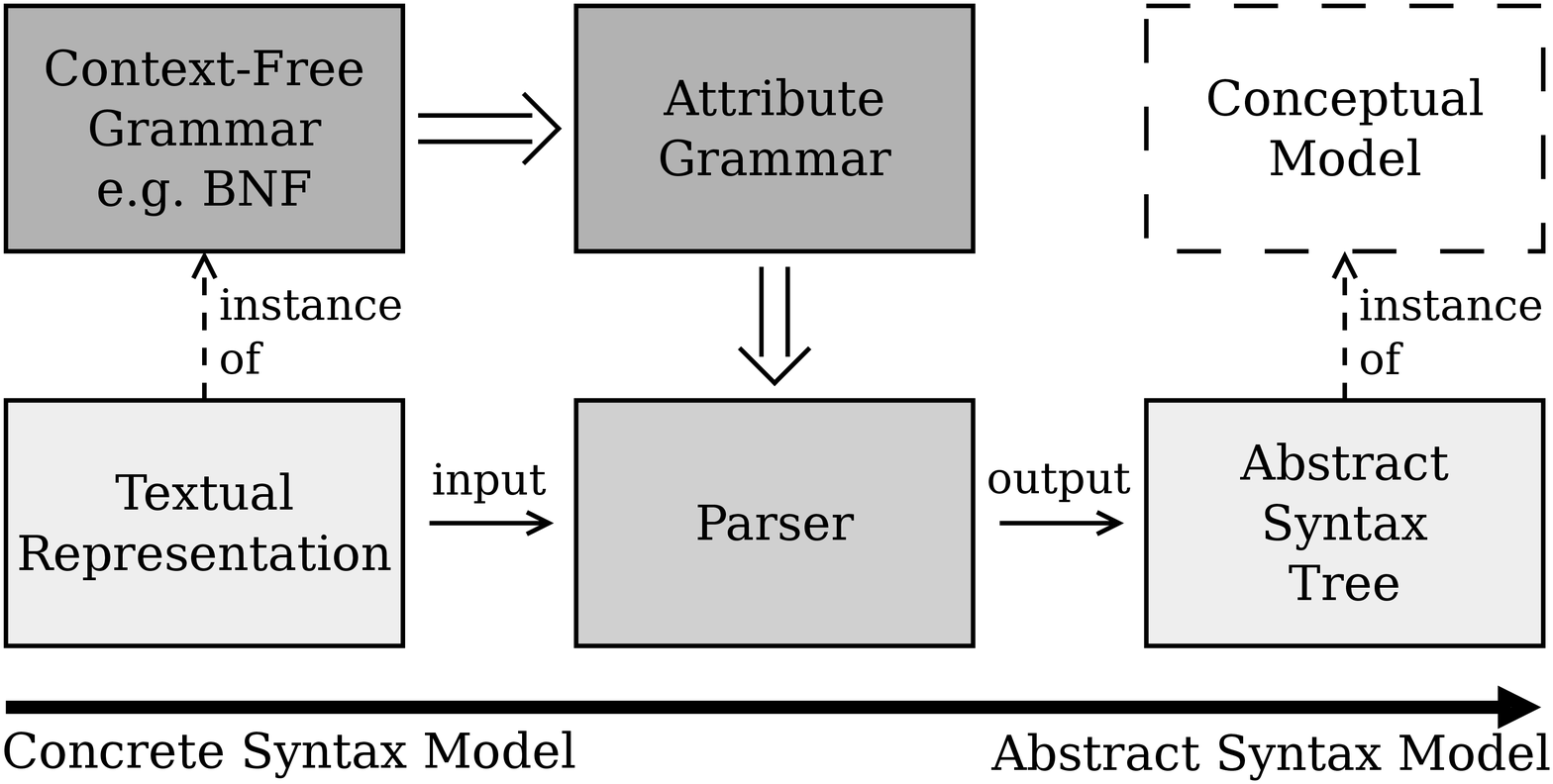}
\caption{Traditional language processing.} \label{fig:traditional}

\vspace{4mm}
\includegraphics[scale=0.225]{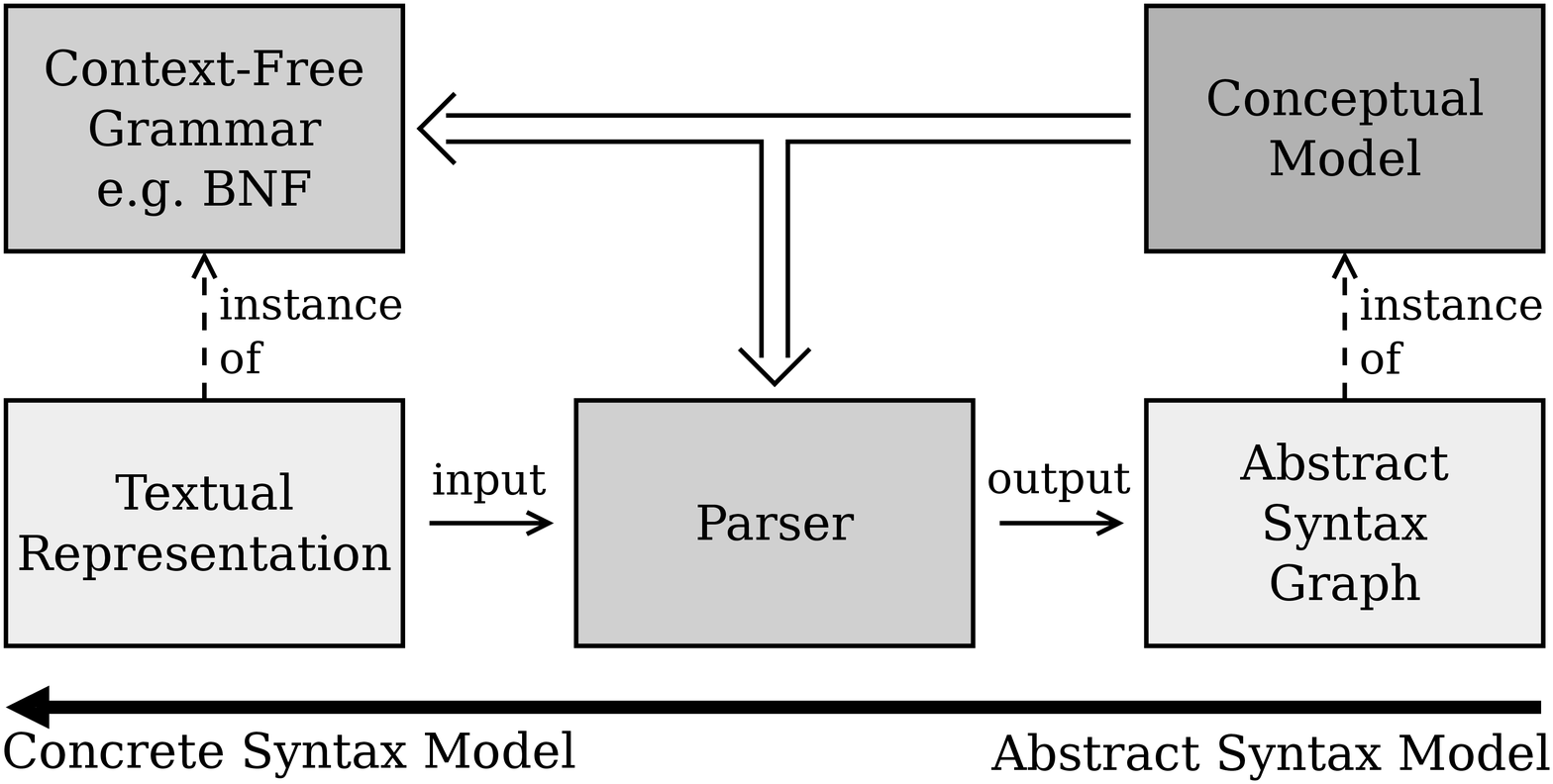}
\caption{Model-based language processing.} \label{fig:ModelCC}
\end{minipage}
\end{figure}

A diagram summarizing the traditional language design process is shown in Figure \ref{fig:traditional}, whereas the corresponding diagram for the model-based approach is shown in Figure \ref{fig:ModelCC}.

It should be noted that ASMs may represent non-tree structures. Hence the use of the `abstract syntax graph' term in Figure \ref{fig:ModelCC}.
When the ASM represents a tree-like structure, a model-based parser generator is equivalent to a traditional grammar-based parser generator in terms of expression power. When the ASM represents non-tree structures, reference resolution techniques can be employed to make model-based parser generators more powerful than grammar-based ones.

ModelCC \cite{Quesada2011c,Quesada2012k} is a parser generator that supports a model-based approach to the design of language processing systems.
Its starting ASM is created by defining classes that represent language elements and establishing relationships among those elements. Once the ASM is established, constraints can be imposed over language elements and their relationships as annotations in order to produce the desired ASM-CSM mappings.

The ASM is built on top of basic language elements, which can be viewed as the tokens in the model-driven specification of a language. ModelCC provides the necessary mechanisms to combine those basic elements into more complex language constructs, which correspond to the use of concatenation, selection, and repetition in the syntax-driven specification of languages.

\section{ModelCC Domain-Specific Language for ASM-CSM mappings} \label{sec:syntaxmodelconstraintspecificationlanguage}

\begin{table*}[tb]
\begin{center}

\setlength{\tabcolsep}{5pt}
\begin{tabular}{ l  l  l } \hline

Constraints on... & Annotation & Function \\ \hline

\multirow{2}{*}{Patterns}
& @Pattern & Pattern definition of basic language elements. \\
& @Value & Field where the matched input will be stored. \\ \hline

\multirow{3}{*}{Delimiters}
& @Prefix & Element prefix(es). \\
& @Suffix & Element suffix(es). \\
& @Separator & Element separator(s). \\ \hline

\multirow{3}{*}{Cardinality}
& @Optional & Optional elements.\\
& @Minimum & Minimum element multiplicity.\\
& @Maximum & Maximum element multiplicity.\\ \hline

\multirow{3}{50pt}{Evaluation order}
& @Associativity & Element associativity (e.g. left-to-right). \\
& @Composition & Eager or lazy policy for nested composites. \\
& @Priority & Element precedence. \\ \hline

\multirow{2}{*}{References}
& @ID & Identifier of a language element. \\
& @Reference & Reference to a language element. \\ \hline

\end{tabular}
\end{center}
\caption{Summary of the basic metadata annotations supported by ModelCC.} \label{fig:tablesummary}
\end{table*}

\noindent In ModelCC, the constraints imposed over ASMs to define a particular ASM-CSM mapping can be declared as metadata annotations on the model itself. Now supported by all the major programming platforms, metadata annotations are often used in reflective programming and code generation \cite{Fowler2002}. Table \ref{fig:tablesummary} summarizes the set of constraints supported by ModelCC for establishing ASM-CSM mappings between ASMs and their concrete representation in textual CSMs.

However, in order to allow the developer to specify several ASM-CSM mappings, ModelCC also allows the specification of separate sets of constraints by using the ModelCC DSL for ASM-CSM mappings.

Using the ModelCC DSL for ASM-CSM mappings instead of metadata annotations to specify ASM-CSM mappings allows the specification of several ASM-CSM mappings for the same ASM by means of separate constraint specification files. This ultimately allows for the proper model-based decoupling of language specification and language processing.

In this section, we describe the ModelCC DSL for ASM-CSM mappings. We provide the ModelCC implementation of that DSL as an ASM complemented with metadata annotations.
As an example of the usage of this language, we also provide the ModelCC implementation of the DSL for ASM-CSM mappings as an ASM complemented with constraint specification files written in the DSL for ASM-CSM mappings itself.

Subsection \ref{sec:langdesc} outlines the ModelCC DSL for ASM-CSM mappings features.
Subsection \ref{sec:modimp} provides the definition of the ModelCC DSL for ASM-CSM mappings as an ASM complemented with metadata annotations.
Subsection \ref{sec:meta} provides the definition of the ModelCC DSL for ASM-CSM mappings as an ASM complemented with several equivalent constraint specification files.

\subsection{Language Features} \label{sec:langdesc}

\noindent The ModelCC DSL for ASM-CSM mappings supports the following features:

\begin{itemize}
\item The definition of constraints on patterns, delimiters, evaluation order, and references between language elements.
\item The property-like specification of constraints for language elements and language element members.
\item The grammar-like specification of the concrete syntax of language elements by means of a regular-expression-like language.
\end{itemize}

While the semantics of property-like constraint definitions is equivalent to that of metadata annotation constraint definitions, grammar-like constraint specification allows for a more intuitive specification of ASM-CSM mappings.

Grammar-like constraint definitions may be more intuitive to traditional language designers who are familiar with syntax-driven language specification tools.
Such constraint definitions can be redundant with the ASM as, for example, they can also include multiplicity constraints.
ModelCC checks and reports if any syntax implicit in grammar-like constraint definitions conflicts with the language ASM.

Finally, ModelCC checks, reports, and ignores any constraints on language elements on language element members that do not exist.

It should be noted that all the features of the ModelCC DSL for ASM-CSM mappings make of ModelCC a complete model-based language workbench.

\subsection{ModelCC Definition of the DSL for ASM-CSM mappings} \label{sec:modimp}

\noindent The ASM of the ModelCC DSL for ASM-CSM mappings is designed first, and then it is mapped to a CSM by imposing constraints by means of metadata annotations on the model classes.

\begin{figure*}[tb]
\centering
\includegraphics[scale=1.05]{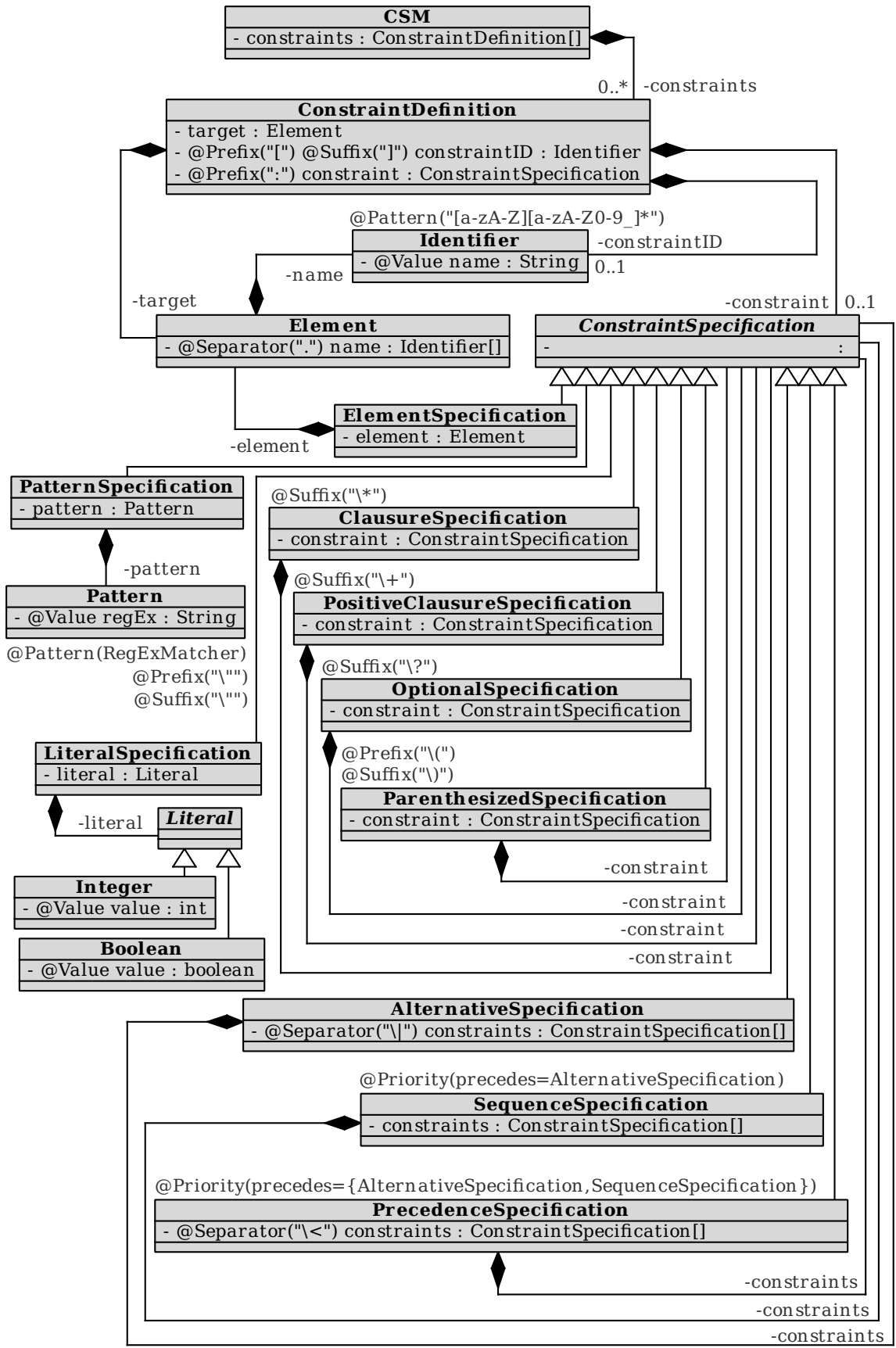}
\caption{Definition of the ModelCC DSL for ASM-CSM mappings in ModelCC.} \label{fig:asmcsm}
\end{figure*}

The resulting model can be processed by ModelCC to generate the corresponding parser.
The ModelCC language model (depicted as an UML class diagram) in Figure \ref{fig:asmcsm} presents the ModelCC DSL for ASM-CSM mappings.

\begin{figure*}[tb]
\centering
\includegraphics[scale=1.05]{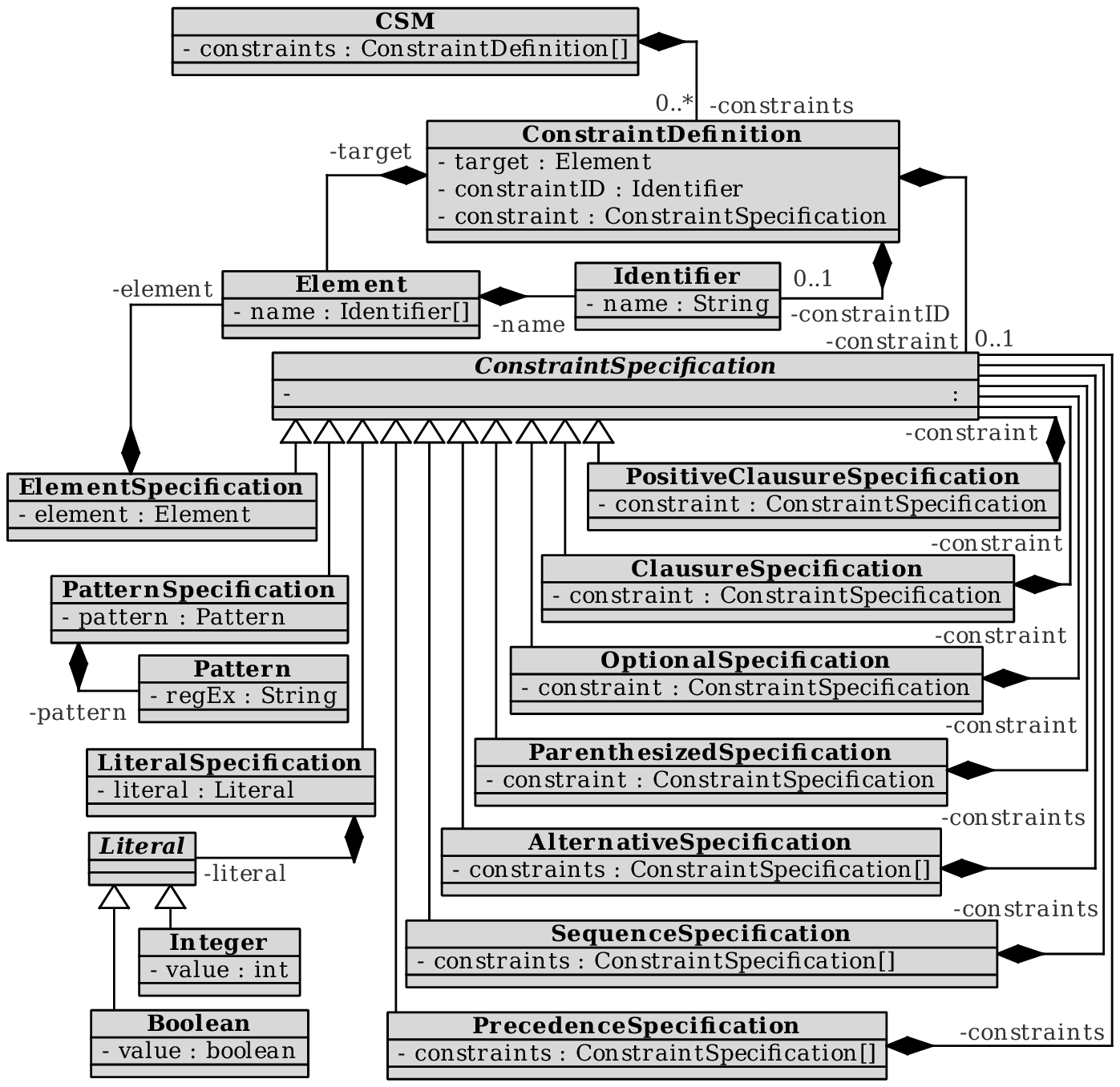}
\caption{Definition of the abstract syntax model of the ModelCC DSL for ASM-CSM mappings in ModelCC.} \label{fig:asm}
\end{figure*}

This Figure highlights two of the reasons why a DSL for ASM-CSM mappings is needed:
\begin{itemize}
\item When metadata annotations are used to define ASM-CSM mappings on top of the ASM, the concrete syntax is interleaved in the abstract syntax model in a way that burdens it, similar to language processing being coupled with language specification in traditional syntax-driven language specification techniques
\item Also, there is no intuitive way to allow the specification of multiple ASM-CSM mappings using metadata annotations.
\end{itemize}

\subsection{Separating the ASM from the CSM} \label{sec:meta}

\noindent Once an initial implementation of the ModelCC DSL for ASM-CSM mappings provides a bootstrap, we provide implementations of the ModelCC DSL for ASM-CSM mappings that consist of an ASM and separate constraint definitions using that language.

The ModelCC language model (depicted as an UML class diagram) in Figure \ref{fig:asm} presents the ASM of the ModelCC DSL for ASM-CSM mappings.

Starting from this ASM, we provide three different ASM-CSM mappings for the language, all of them written in the ModelCC DSL for ASM-CSM mappings itself.
By complementing the ASM with any of the three following ASM-CSM mappings or any other equivalent one, we will obtain the same language as in the previous Section.

\begin{itemize}
\item {\bf Property-like specification} Figure \ref{fig:property} presents a property-like ASM-CSM mapping written in the ModelCC DSL for ASM-CSM mappings.

The property-like specification of ASM-CSM mappings mimics the specification of constraints on ASMs using metadata annotations.
It can be observed that the constraints are specified as properties of language elements.

\begin{figure*}[p]
\centering
\begin{varwidth}{0.99\linewidth}
\begin{verbatim}
ConstraintDefinition.constraintID[prefix] "\["
ConstraintDefinition.constraintID[suffix] "\]"
ConstraintDefinition.constraint[prefix]: ":"
Element.name[separator]: "."
Identifier.name: "[a-zA-Z][a-zA-Z0-9_]*"
ClausureSpecification[suffix]: "\*"
OptionalSpecification[suffix]: "\?"
PositiveClauseSpecification[prefix]: "\+"
ParenthesizedSpecification[prefix]: "\("
ParenthesizedSpecification[suffix]: "\)"
SequenceSpecification[precedes]: AlternationSpecification
                                 PrecedenceSpecification
ConstraintSpecification: SequenceSpecification < PrecedenceSpecification
AlternationSpecification.constraints[separator]: "\|"
PrecedenceSpecification[precedes]: AlternationSpecification
PrecedenceSpecification.constraints[separator]: "\<"
Boolean.value: "true|false"
Integer.value: "[0-9]+"
\end{verbatim}
 \end{varwidth}
\caption{Property-like specification of the mapping from the abstract syntax model to the concrete syntax model of ModelCC DSL for ASM-CSM mappings, written in the ModelCC DSL for ASM-CSM mappings itself.}
\label{fig:property}
\end{figure*}

\begin{figure*}[p]
\centering
\begin{varwidth}{0.99\linewidth}
\begin{verbatim}
ConstraintDefinition: target ("[" constraintID "]")? (":" constraint)?
Element: name ("." name)*
Identifier.name: "[a-zA-Z][a-zA-Z0-9_]*"
ClausureSpecification: constraint "\*"
OptionalSpecification: constraint "\?"
PositiveClauseSpecification: constraint "\+"
ParenthesizedSpecification: "\(" constraint "\)"
ConstraintSpecification: SequenceSpecification < PrecedenceSpecification
                         < AlternationSpecification 
AlternationSpecification: constraints ("\|" constraints)*
PrecedenceSpecification: constraints ("\<" constraints)*
Boolean.value: "true|false"
Integer.value: "[0-9]+"
\end{verbatim}
 \end{varwidth}
\caption{Grammar-like specification of the mapping from the abstract syntax model to the concrete syntax model of ModelCC DSL for ASM-CSM mappings, written in the ModelCC DSL for ASM-CSM mappings itself.}
\label{fig:grammar}
\end{figure*}

\begin{figure*}[p]
\centering
\begin{varwidth}{0.99\linewidth}
\begin{verbatim}
ConstraintDefinition: "[" constraintID "]"
ConstraintDefinition: ":" constraint
Element.name[separator]: "."
Identifier.name: "[a-zA-Z][a-zA-Z0-9_]*"
ClausureSpecification: constraint "\*"
OptionalSpecification: constraint "\?"
PositiveClauseSpecification: constraint "\+"
ParenthesizedSpecification: "\(" constraint "\)"
ConstraintSpecification: SequenceSpecification < PrecedenceSpecification
                         < AlternationSpecification 
AlternationSpecification.constraints[separator]: "\|"
PrecedenceSpecification.constraints[separator]: "\<"
Boolean.value: "true|false"
Integer.value: "[0-9]+"
\end{verbatim}
 \end{varwidth}
\caption{Mixed specification of the mapping from the abstract syntax model to the concrete syntax model of ModelCC DSL for ASM-CSM mappings, written in the ModelCC DSL for ASM-CSM mappings itself.}
\label{fig:mixed}
\end{figure*}
\item {\bf Grammar-like specification} Figure \ref{fig:grammar} presents a grammar-like ASM-CSM mapping written in the ModelCC DSL for ASM-CSM mappings.

Some of the advantages of grammar-like mappings can be observed in the specification of the \emph{ConstraintDefinition} language element constraints.
A single constraint specification can include prefix constraints, suffix constraints, and language element member order constraints.
Also, the specification of the \emph{ConstraintDefinition} language element constraints includes two multiplicity constraints (optionality, represented by the regex-like ``?'' operator) that are redundant with the ASM.
ModelCC will check these multiplicity constraints for consistency with the ASM, and will report any conflict in parser generation time.

Another illustrative case of grammar-like mappings can be observed in the specification of the \emph{Element} language element constraints.
Although its member \emph{name} is defined as a list in the ASM, the grammar-like constraint specification uses a classical explicit-list specification to specify the separator for list members.

\item {\bf Mixed specification} Figure \ref{fig:mixed} presents another example of an ASM-CSM mapping written in the ModelCC DSL for ASM-CSM mappings. In this case, some constraints are specified grammar-like and some constraints are specified property-like.

In this case, separators in lists are specified using property-like constraint definitions, which may seem more intuitive to some language designers.

It should be noted that constraint definitions differ from grammar rules in that several of them can be specified for separate members of the same language element, as can be observed in the \emph{ConstraintDefinition} language element.
\end{itemize}

Finally, it should be noted that ASMs that are complemented with metadata annotations can be complemented with files written in the ModelCC DSL for ASM-CSM mappings.
Such files could redefine constraints that are specified in the original annotated ASM.
Therefore, metadata annotation constraints would represent default values that would apply, unless otherwise specified, to all the ASM-CSM mappings of a language.

\section{Conclusions and Future Work} \label{sec:conclusionsandfuturework}

\noindent ModelCC is a model-based parser generator that allows using metadata annotations or a domain-specific language to specify abstract syntax model to concrete syntax model mappings.

In this paper, we have proposed and described the ModelCC domain-specific language for abstract syntax model to concrete syntax model mappings (ModelCC DSL for ASM-CSM mappings).

The ModelCC DSL for ASM-CSM mappings allows the specification of separate abstract syntax model to concrete syntax model mappings and, therefore, effectively decouples abstract syntax models from concrete syntax models.

As an example, we have specified the ModelCC DSL for ASM-CSM mappings as an ASM and several equivalent ASM-CSM mappings written in the ModelCC DSL for ASM-CSM mappings language itself.

In the future, we plan to apply model-based language specification techniques to problems such as data integration and natural language processing.
We also plan to incorporate different reference resolution techniques to ModelCC.

\vspace{1mm}
\bibliographystyle{plain}
\bibliography{doc}

\begin{thebibliography}{1}

\bibitem{Aho2006}
Alfred~V. Aho, Monica~S. Lam, Ravi Sethi, and Jeffrey~D. Ullman.
\newblock {\em Compilers: Principles, Techniques, and Tools}.
\newblock Addison Wesley, 2nd edition, 2006.

\bibitem{Fowler2002}
Martin Fowler.
\newblock Using metadata.
\newblock {\em IEEE Software}, 19(6):13--17, November 2002.

\bibitem{Kats2010}
Lennart C.~L. Kats, Eelco Visser, and Guido Wachsmuth.
\newblock Pure and declarative syntax definition: Paradise lost and regained.
\newblock In {\em Proceedings of the ACM International Conference on
  Object-Oriented Programming Systems, Languages, and Applications
  (OOPSLA'10)}, pages 918--932, 2010.

\bibitem{Kleppe2007}
Anneke Kleppe.
\newblock Towards the generation of a text-based {IDE} from a language
  metamodel.
\newblock volume 4530 of {\em Lecture Notes in Computer Science}, pages
  114--129, 2007.

\bibitem{Quesada2012k}
Luis Quesada.
\newblock A model-driven parser generator with reference resolution support.
\newblock In {\em Proceedings of the 27th IEEE/ACM International Conference on
  Automated Software Engineering}, pages 394--397, 2012.

\bibitem{Quesada2011c}
Luis Quesada, Fernando Berzal, and Juan-Carlos Cubero.
\newblock A language specification tool for model-based parsing.
\newblock In {\em Proceedings of the 12th International Conference on
  Intelligent Data Engineering and Automated Learning. Lecture Notes in
  Computer Science}, volume 6936, pages 50--57, 2011.

\end{thebibliography}

\end{document}